\documentclass{article}
\usepackage{geometry}
\usepackage{graphicx}

\title{	Effects of curvature on hydrothermal waves instability 
of radial thermocapillary flows \\
	Effets de courbure sur l'instabilit\'e en ondes hydrothermales
d'un \'ecoulement thermocapillaire radial}

\author{Nicolas Garnier$^\dag$ and Christiane Normand$^\ddag$\\
$\dag$ Service de Physique de l'Etat Condens\'e, 
$\ddag$ Service de Physique Th\'eorique\\
CEA/Saclay 91191 Gif-sur-Yvette CEDEX, France}

\date{}

\begin{document}

\maketitle

\begin{tabbing}
published reference: Comptes-rendus de l'Academie des Sciences 2001 {\bf 2} (8) pp1227-1233 (2001). \\
Corresponding author: Christiane Normand
e-mail: {\tt normand@spht.saclay.cea.fr}\\
\end{tabbing}

\section*{Abstract}

The stability of a thermocapillary flow in an extended cylindrical
geometry is analyzed. This flow occurs in a thin liquid layer with a
disk shape when a radial temperature gradient is applied along the
horizontal free surface. Besides the aspect ratio, a second parameter
related to the local curvature is introduced to describe completely the
geometrical effects. We recover classical hydrothermal waves as
predicted by Smith and Davis~\cite{SD83} but the properties of these
waves are shown to evolve with the curvature parameter, thus leading to
a non uniform pattern over the cell. Moreover, it is shown that the
problem is not invariant with respect to the exchange of the hot and
cold sides.

\vspace{.5cm}
Keywords: Hydrodynamic instabilities, hydrothermal waves, thermocapillary flow

\section*{R\'esum\'e}

Nous \'etudions la stabilit\'e lin\'eaire d'un \'ecoulement
thermocapillaire en g\'eom\'etrie cylindrique \'e\-ten\-due. Un tel
\'ecoulement est produit dans un disque de fluide dont la surface libre
horizontale est soumise \`a un gradient de temp\'erature purement
radial. Outre le rapport d'aspect, un second param\`etre li\'e \`a la
courbure locale est introduit pour caract\'eriser la g\'eom\'etrie du
probl\`eme. L'instabilit\'e en ondes hydrothermales pr\'edite
par~\cite{SD83} est retrouv\'ee mais les propri\'et\'es des ondes sont
alt\'er\'ees par la courbure locale ce qui explique l'existence de
structures non uniformes. La dissym\'etrie du probl\`eme vis-\`a-vis
d'une inversion du gradient de temp\'erature est aussi mise en
\'evidence.

\vspace{0.5cm}
Mots cl\'es: Instabilit\'es hydrodynamiques, ondes hydrothermales, \'ecoulement thermocapillaire
\newpage

\section{Sketch of the problem}

When a horizontal temperature gradient is imposed over a thin fluid
layer with a free surface, a basic flow is present due to the Marangoni
effect, {\it i.e.}, surface tension dependence on the temperature. When
the temperature difference is small, the characteristic velocity of the
flow is proportional to the thermal constraint and the flow is
stationary. When the temperature difference is increased above a
threshold value, the basic flow is unstable to propagating waves called
hydrothermal waves and first predicted by Smith and Davis~\cite{SD83}
for a zero gravity environment. 

The stability analysis of the basic flow has been addressed by many
authors in Cartesian coordinates \cite{Mercier96}, corresponding to
experiments in rectangular cavities~\cite{Riley98}, but a very few
theoretical studies were devoted to the cylindrical
geometry~\cite{Vrane96} though experiments are often conducted in this
configuration~\cite{Schwabe92}. The present paper aims to fill the gap
and investigates the effect of curvature on the critical properties of
hydrothermal waves.

\begin{figure}[htbp]
\begin{center} \includegraphics[scale=0.5]{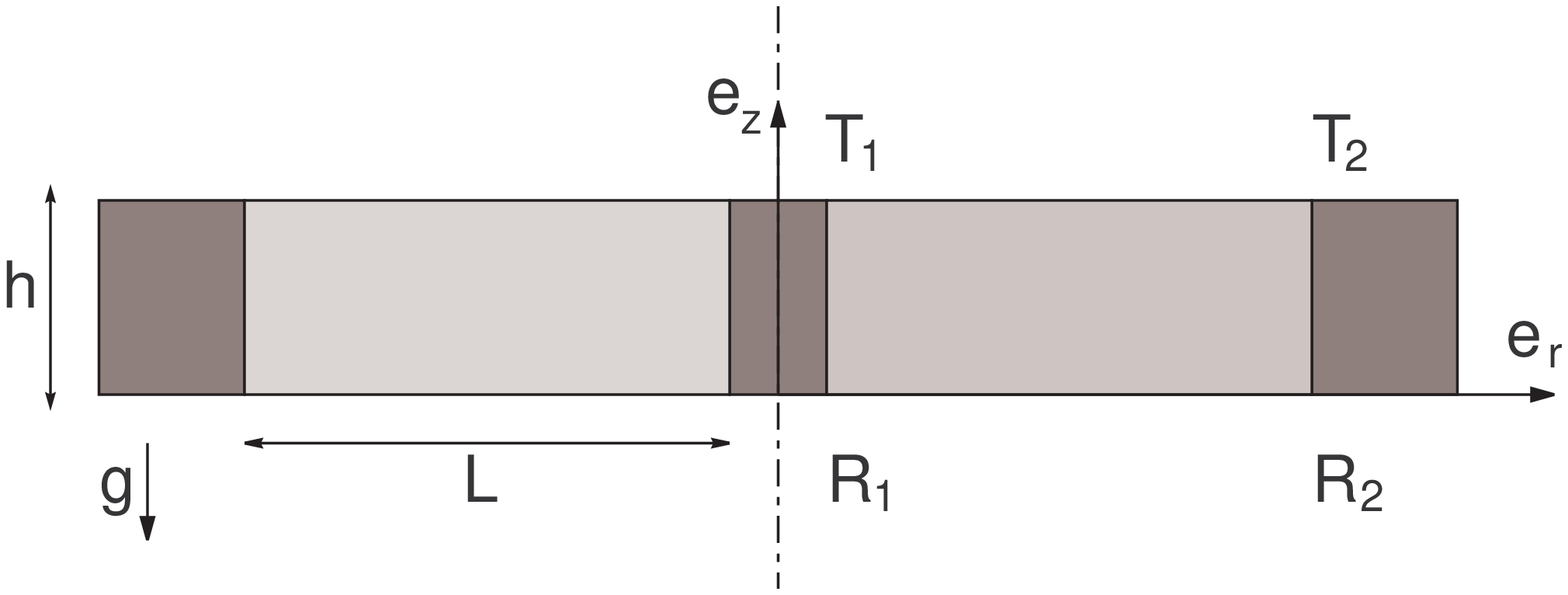} \end{center}
\caption{Section of the cylindrical geometry and associated notations.}
\label{fig:geometry}
\end{figure}

The section of a typical cell in a vertical diametrical plane is depicted
on Fig.~\ref{fig:geometry}. We note $h$ the depth of the fluid layer,
and $R_1, R_2=R_1+L$ the radii of the internal and external cylindrical
boundaries that confine the fluid in the horizontal direction. The
horizontal temperature gradient is supposed to be applied as a
temperature difference between the inner and the outer boundaries,
considered as isothermal with temperature $T_1$ and $T_2$ respectively.
The bottom of the cell is considered as perfectly
conducting~\cite{garnier00} and a mixed thermal boundary condition is used for
the top free surface~\cite{Mercier96}. The fluid is assumed to be
incompressible with kinematic viscosity $\nu$, thermal diffusivity
$\kappa$, density $\rho$, thermal expansion coefficient $\alpha$. The
surface tension $\sigma$ of the interface varies with the temperature
$T$ of the liquid and its derivative ${\partial \sigma / \partial
T}=-\gamma$, is considered as constant and negative ($\gamma>0$).

\section{Formulation of the basic flow}

To fully characterize the geometrical configuration, two non-dimensional parameters are
required. First,
the aspect ratio $\varepsilon = h/L$ which is supposed to be small in the present 
extended geometry ($\varepsilon<0.03$). Second, the local curvature proportional to the inverse 
of the radius will be represented in non dimensional form by $L/r$.
Its maximum value: $\phi = L/R_1$, is taken as
the additional parameter characteristic of the cylindrical geometry as a whole. 
 When the change of 
variable: $r= R_1 + L X$, is made the local curvature is expressed as
$$
\Phi(X) = \frac{\phi}{1+\phi X}=\frac{L}{r}
$$
The basic flow is governed by the Navier-Stokes equations in the Boussinesq 
approximation together with the heat equation. Owing to rotational invariance
 of the system about the vertical axis
$e_z$ of the cell, the basic flow has no azimuthal velocity component.
The governing equations are written in non-dimensional form by taking
the following scales~: $h$ for the vertical coordinate $z$,  $\nu/h$ for vertical velocity $w$
and radial velocity $u$, $h2/\nu$ for the time $t$, $\nu2 \rho_0 L / h3 $ 
for the pressure $P$ and $T_2-T_1$
for the temperature difference $T-T_1$. 

The boundary conditions represent a conducting rigid bottom at $z=0$ 
and a mixed thermal condition on the top free surface at $z=1$~:
\begin{eqnarray}
u=w=0, \quad \mbox{and} \quad T=T_c(X)
	& \mbox{on} \quad z=0, \label{b0}\\
w=0,   \quad \left({\partial u \over \partial z} + \varepsilon {\partial w \over \partial X}\right)
		+ {\rm Re}{\partial T \over \partial X}=0,
       \quad \mbox{and} \quad {\partial T \over \partial z} + \mathrm{Bi}(T-T_c(X))=0, 
	& \mbox{on} \quad z=1, \label{b1}
\end{eqnarray}
where $T_{\rm c}(X) = \ln (1+\phi X)/ \ln (1+\phi)$ is the conductive
temperature profile in the cylindrical geometry and $\mathrm{Bi}$ is the
Biot number. The Reynolds number ${\rm Re}=\gamma(T_2-T_1)h2/\nu2 \rho_0 L$ is
related to the Marangoni number by ${\rm Ma}={\rm Re}{\rm Pr}$ where ${\rm Pr}=\nu/\kappa$ is
the Prandtl number of the fluid.

Far from the lateral boundaries, the basic state $(u_0,0,w_0,T_0)$ is determined as an
expansion into
the small parameter $\varepsilon$, as performed by~\cite{Vrane96}. Up to first
order in $\varepsilon$, we obtain:

\begin{equation}
\left\{ \begin{array}{ll}
u_0(X,z) &= {\rm Gr} \Phi(X) \bar{u}_0(W,z) + \varepsilon {\rm Gr}^2 \Phi(X)3 \bar{u}_1(W,z)+{\cal O}({\varepsilon}^2)\\
w_0(X,z) &= {\cal O}({\varepsilon}^2)\\
T_0(X,z) &= T_{\rm c}(X) + \varepsilon {\rm Ra} \Phi(X)2 \bar{T}_0(W,z)+{\cal O}({\varepsilon}^2)
\end{array} \right.
\label{basic}
\end{equation}
with the Grashof number: 
$$ {\rm Gr}= {\alpha g (T_2-T_1)h4 \over L\nu2}$$
and the Rayleigh number ${\rm Ra}={\rm Gr} {\rm Pr}$. The parameter $W={\rm Re}/{\rm Gr}$, which
behaves like $h^{-2}$ is the inverse of the Bond number, it measures the
relative importance of thermocapillary and thermogravity effects. The
functions $\bar{u}_0(W,z)$ and $\bar{T}_0(W,z)$ are polynomials in $z$
similar to those that describe the velocity and temperature profiles in
rectangular geometry~\cite{Mercier96} and $\bar{u}_1(W,z)$ is given in \cite{nico}. 
The main difference with the
rectangular geometry is the non uniformity of the flow along the radial
direction which appears through the slowly varying function $\Phi(X)$.
In the forthcoming stability analysis this will prevent decomposition of
the disturbances in Fourier modes in the radial direction. To bypass
this difficulty we shall perform a local stability analysis and
introduce the local Grashof number ${\rm Gr}_X={\rm Gr} \Phi(X)$ as the governing
parameter and similarly ${\rm Ra}_X={\rm Ra} \Phi(X)$. In the following we shall assume
that ${\rm Gr}_X={\cal O}(1)$. Thus, the second term in the expansion of $u_0$ and $T_0$
is of the order $\varepsilon \Phi(X)=\Gamma(X)$ which is a small quantity
only near the outer cylinder where $\Gamma(1) \approx \varepsilon$. 
Near the inner cylinder, $\Gamma(X)$ takes values as large as $\Gamma(0)=0.475$ \cite{nico}
and cannot be neglected. It will be shown in
the next section that the perturbations equations depend on $X$
exclusively through ${\rm Gr}_X$ and $\Gamma(X)$.
 
\section{Linear stability analysis}

The basic state~(\ref{basic}) is perturbed by the superposition of
three-dimensional disturbances $\vec{v}$, for the velocity and $T$, for
the temperature such as 
\begin{equation} 
\left\{ \vec{v}(r,\theta,z,t), T(r,\theta, z ,t)\right\}= 
 {\exp i(m \theta+\omega t)} \left\{\hat{u}(r,z),\hat{v}(r,z),\hat{w}(r,z),\hat{T}(r,z)\right\}
\label{pert}
\end{equation} 
where $m$ is the azimuthal wave number and $\omega$ the
frequency. The disturbances are assumed to evolve in the horizontal
plane on a smaller scale than the basic flow. Thus, the fast variable
$x=\varepsilon^{-1}X$ is introduced and the azimuthal wave number $m$ is
replaced by the local wave number $\beta=\Gamma(X) m$. If the
stability analysis is performed while $\Gamma$ is kept constant, the
disturbances can be sought as periodic functions of $x$ such as for
example $\hat{u}(r,z)=u(z)\exp i\alpha x$. The linearized evolution
equations for the perturbations reduce to a set of coupled differential
equations 
\begin{eqnarray} 
{\rm Gr}_X \left[( (i \alpha - \Gamma)\bar{u}_0 +\Gamma {\rm Gr}_X (i \alpha - 3\Gamma)\bar{u}_1)\right. u \nonumber\\
\left. \mbox{}+(\bar{u}_0' + \Gamma{\rm Gr}_X \bar{u}_1') w \right]  &=&
-i\alpha p +(L-i \omega-\Gamma2)u-2i \beta \Gamma v \label{pu} \\
{\rm Gr}_X (i\alpha+\Gamma) (\bar{u}_0 +\Gamma {\rm Gr}_X \bar{u}_1) v &=&
-i \beta p +(L-i\omega-\Gamma2)v + 2i \beta \Gamma u, \label{pv}\\
i\alpha {\rm Gr}_X (\bar{u}_0 +\Gamma {\rm Gr}_X \bar{u}_1) w &=&
-{\partial p \over \partial z} +(L-i \omega) w + T, \label{pw}\\ 
{\rm Ra}_X\left[i\alpha (\bar{u}_0 +\Gamma {\rm Gr}_X \bar{u}_1) T+ (1-2\Gamma {\rm Ra}_X \bar{T}_0)u  \right. \nonumber\\ 
\left.\mbox{}+ {\rm Ra}_X \bar{T}_0' w\right]& = &(L-iPr \omega)T \label{pT}
\end{eqnarray} 
with $L=\partial2 /\partial z2-(\alpha2 +\beta2)+i\alpha \Gamma$. The incompressibility 
condition reads
\begin{equation} 
(i\alpha +\Gamma)u+i\beta v +{\partial w \over \partial z} =0, \label{divu} 
\end{equation} 
The boundary conditions are 
\begin{eqnarray} u=v=w=T=0, \quad \mbox{at} \quad z=0,\label{bc0}\\
 w=0, \quad \partial_z T+ Bi T=0, \quad
\partial_z u + i \alpha W T=0, \quad \partial_z v +i \beta W T=0. \quad
\mbox{at} \quad z=1. \label{bc1} 
\end{eqnarray} 
When $\Gamma=0$, the permutation of the cold and hot sides has no
consequence on the instability pattern. Indeed, the transformation ($Gr,
\alpha, u \to -Gr, -\alpha, -u$) leaves the above system invariant. This
is no longer true when $\Gamma \ne 0$, in which case the sign of
$\Gamma$ needs also to be changed jointly for the system to be
invariant. For fixed values of the curvature parameter $\Gamma$, the
differential system is solved by the same method as in \cite{Mercier96}
and critical values of the Grashof number, frequency and wave vector are
found as functions of the curvature parameter $\Gamma$.

\begin{figure}[htbp]
\begin{center} \includegraphics[scale=0.8]{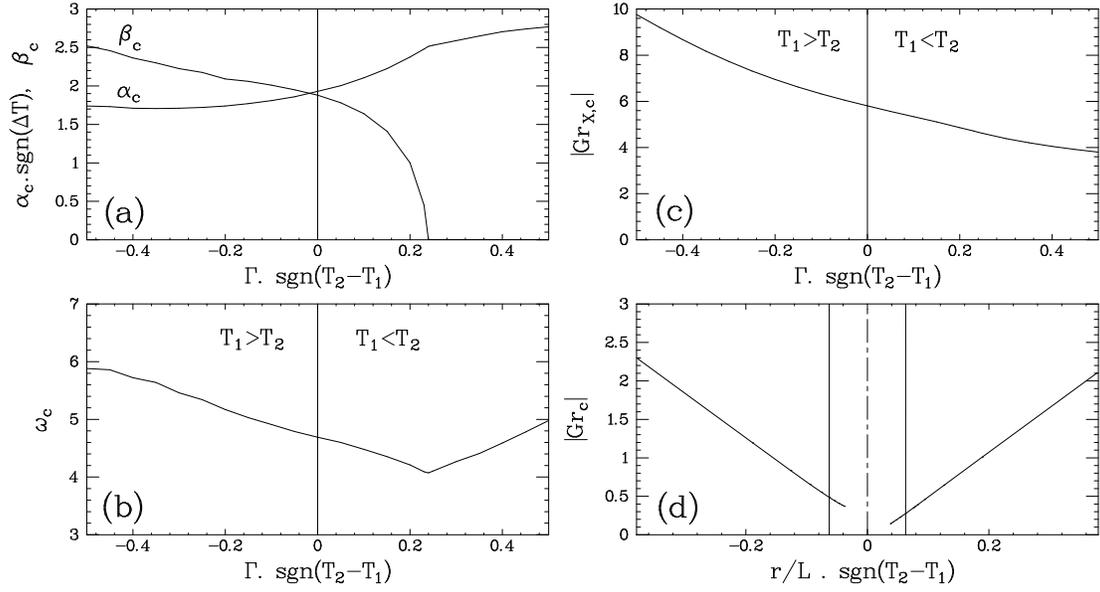}
\end{center}
\caption{Critical values of (a) the wavenumbers $(\alpha, \beta)$, 
(b) the frequency $\omega$ and (c) the local Grashof number $Gr_X$ as 
functions of the signed curvature. (d): Grashof number $Gr$ versus
signed radial position.}
\label{fig:results}
\end{figure}

The results presented on Fig.~\ref{fig:results} are obtained for values
of the physical and geometrical parameters $(Pr=10, Bi=1, \phi=16, W=6)$
corresponding to experiments described in~\cite{garnier00}. On this
figure, we have used the signed curvature ${\rm{sgn}}(T_2-T_1)
\Gamma(X)$ and we can then observe the smooth behavior of all the
quantities when crossing the zero value corresponding to the rectangular
case. Results obtained for $W=2.4$ \cite{nico}, which are not reported
here, confirm the behavior of the critical quantities shown in Fig. 2.

\section{Discussion}

From Fig.~2c showing the critical value of the local Grashof number as
function of the local curvature we deduced Fig.~2d showing the critical
Grashof number $Gr_c$ versus the radial position, which is more convenient for
comparison with experimental results. Particularly interesting is the
decrease of the critical Grashof number when $r \to 0$. This means that
hydrothermal waves will first appear near the center of the cylindrical
cell, i.e., in region of higher curvature. An increase in the curvature
also leads to a variation of the frequency (Fig.~2b) and over all to a
significant variation of the wave vector orientation (Fig.~2a). 

When $T_2>T_1$, the azimuthal wave number $\beta$ vanishes for a value of
the curvature $\Gamma^*=0.24$, which means that in cells with enough
curvature, hydrothermal waves will be not only localized near the
center, but they will propagate in the radial direction at onset. This
fact was observed experimentally in~\cite{garnier00} where shadowgraphic
pictures of the instability pattern clearly show that at onset the
spatial structure is made of concentric circles called targets, which
are localized near the cold center of the cell. Slightly above onset the
pattern spreads toward the outer side of the cell and deforms
progressively to give rise to spirals that fill more and more space as
the supercritical region extends (Fig.~2d), in agreement with
experimental observations (Fig. 6,7 in~\cite{garnier00}).

When the center of the cell is the hot side ($T_1>T_2$), the situation
is quite different and there is experimental evidence that the
instability pattern has always a non vanishing azimuthal wave number in
agreement with our theoretical results. For a small height of fluid the
pattern is made of spirals localized near the center of the cell \cite{nico}. For a
higher height of fluid, an hexagonal structure is shown at onset \cite{nico}.

\section{Conclusion}

The flow which appears in a thin fluid layer confined between two
differentially heated cylinders was considered. Far from the side walls,
the basic state is modeled by a slowly varying flow along the radial
direction. A local stability analysis is performed leading to the
critical values of the Grashof number, wave numbers and frequency as
functions of the local curvature. As a result, the instability is
predicted to appear first near the inner cylinder and the values of the
corresponding wave numbers are quite different whether the inner cylinder
is the hot or the cold side. This is in agreement with experiments where
spiraling waves have been observed near the hot inner side \cite{nico}  while
pulsating targets appears when the inner cylinder is the cold side \cite{garnier00}.

\newpage
\setcounter{section}{0}

\section*{Version francaise abr\'eg\'ee}

\section{Introduction}

Lorsque la surface libre d'une couche de fluide de faible \'epaisseur
est soumise \`a un gradient thermique horizontal il appara\^{\i}t un
\'ecoulement d\^u \`a l'effet Marangoni qui r\'esulte de la variation de
la tension superficielle avec la temp\'erature. Cet \'ecoulement qui est
stationnaire pour de faibles valeurs du gradient thermique, se
d\'estabilise au-del\`a d'un seuil \`a partir duquel il appara\^{\i}t
des ondes propagatives hydrothermales selon la terminologie de Smith et
Davis~\cite{SD83}.
 
Les analyses de stabilit\'e \cite{Mercier96} de ce type d'\'ecoulement
\'etaient jusqu'\`a pr\'esent surtout adapt\'ees aux r\'ealisations
exp\'erimentales en cellules rectangulaires \cite{Riley98}. Peu
d'\'etudes th\'eoriques traitent de la g\'eom\'etrie cylindrique
\cite{Vrane96} alors que de nombreuses exp\'eriences sont r\'ealis\'ees
dans cette configuration \cite{Schwabe92}, \cite{garnier00},
\cite{nico}. Pour y rem\'edier, nous analysons l'effet de la courbure
sur les propri\'et\'es critiques des ondes hydrothermales.

Un sch\'ema du syst\`eme consid\'er\'e est r\'epr\'esent\'e en figure 1.
Une couche de fluide d'\'epaisseur $h$ est confin\'ee lat\'eralement par
deux cylindres de rayons $R_1$ et $R_2=R_1+L$ maintenus aux
temp\'eratures $T_1$ et $T_2$ respectivement. Le fluide est
incompressible, de viscosit\'e $\nu$, diffusivit\'e thermique $\kappa$,
densit\'e $\rho$, coefficient de dilatation thermique $\alpha$. La
tension superficielle $\sigma$ de l'interface varie avec la
temp\'erature $T$ du fluide et ${\partial \sigma / \partial T}=-\gamma$,
avec $\gamma$ constant et positif.

\section{Etat de base}

Deux param\`etres sans dimension sont n\'ecessaires pour caract\'eriser
compl\'etement la g\'eom\'etrie. D'une part, le rapport d'aspect
$\varepsilon = h/L$, et d'autre part le maximum de la courbure locale
$\phi=L/R_1$. Effectuant le changement de variable radiale $r=R_1+LX$,
la courbure locale $L/r$, est repr\'esent\'ee par la fonction
$\Phi(X)=\phi/(1+\phi X)$.

La vitesse et la temp\'erature du fluide sont r\'egies par les
\'equations de Navier-Stokes dans l'approximation de Boussinesq et par
l'\'equation de la chaleur. Un choix d'\'echelles appropri\'e permet
d'\'ecrire ces \'equations pour des grandeurs sans dimension faisant
ainsi appara\^{\i}tre le nombre de Reynolds $Re=\gamma(T_2-T_1)h2/\nu2
\rho_0 L$ et le nombre de Grashof $Gr=\alpha g (T_2-T_1)h4 / L\nu2$ dont
on notera le rapport $W=Re/Gr$. Apr\`es multiplication par le nombre de
Prandtl $Pr=\nu/\kappa$ du fluide on obtient respectivement le nombre de
Marangoni $Ma=RePr$ et le nombre de Rayleigh $Ra=Gr Pr$. Les conditions
aux limites qui s'expriment par les relations (\ref{b0}) et (\ref{b1})
repr\'esentent une paroi inf\'erieure rigide et conductrice et une
surface sup\'erieure libre et plane o\`u les \'echanges thermiques sont
d\'ecrits par l'interm\'ediaire d'un nombre de Biot $Bi$.

Loin des parois cylindriques qui confinent le fluide lat\'eralement, la
vitessse et la temp\'erature dans l'\'etat de base ($u_0,0,w_0,T_0$)
sont d\'etermin\'ees par un d\'eveloppement en puissance de
$\varepsilon$ (\ref{basic}). La variation lente de l'\'ecoulement de
base dans la direction radiale est report\'ee dans la d\'efinition du
nombre de Grashof local $Gr_X=Gr\Phi(X)$, permettant ainsi d'effectuer
une analyse de stabilit\'e locale o\`u $Gr_X$ repr\'esente le
param\`etre de contr\^ole et o\`u intervient le param\`etre
g\'eom\'etrique $\Gamma(X)=\varepsilon \Phi(X)$ mesurant la courbure
locale.
 
\section{Analyse de stabilit\'e lin\'eaire}

On superpose \`a l'\'etat de base des perturbations tri-dimensionnelles
{$\vec v$} pour la vitesse et $T$ pour la temp\'erature repr\'esent\'ees
par des modes (\ref{pert}) de nombre d'onde azimutal $m$ et de
fr\'equence $\omega$. Nous consid\'erons des perturbations qui
\'evoluent dans le plan horizontal sur une \'echelle de longueur plus
courte que celle de l'\'ecoulement de base, ce qui conduit \`a
introduire la variable $x=\varepsilon^{-1}X$ et \`a remplacer le nombre
d'onde azimutal $m$ par $\beta=\Gamma(X)m$. Pour une valeur fix\'ee de
$\Gamma$, les perturbations admettent comme solutions des fonctions
p\'eriodiques en $x$, de nombre d'onde radial $\alpha$. Apr\`es
lin\'earisation, les \'equations d'\'evolution pour les perturbations se
r\'eduisent au syst\`eme d'\'equations diff\'erentielles
(\ref{pu})-(\ref{pT}), avec la condition d'incompressibilit\'e
(\ref{divu}) et les conditions aux limites associ\'ees (\ref{bc0}) and
(\ref{bc1}).
 
En l'absence de courbure, $\Gamma=0$, la permutation des bords chaud et
froid n'a pas de cons\'equence majeure puisque le syst\`eme
(\ref{pu})-(\ref{pT}) est invariant dans la transformation ($Gr, \alpha,
u \to -Gr, -\alpha, -u$). Il n'en va plus de m\^eme lorsque $\Gamma \ne
0$, auquel cas il faut inverser conjointement $\Gamma$ pour conserver
l'invariance du syst\`eme. Apr\`es r\'esolution num\'erique du syt\`eme
(\ref{pu})-(\ref{pT}) pour les valeurs des param\`etres $(Pr=10, Bi=1,
\phi=16, W=6)$, on pr\'esente en figure 2 les valeurs critiques du
nombre de Grashof, de la fr\'equence et des nombres d'onde en fonction
de la courbure.

\section{Discussion}

La variation de la valeur critique du nombre de Grashof en fonction de
la position radiale est repr\'esent\'ee sur la figure 2d. On remarque une
d\'ecroissance de $Gr_c$ lorsque $r \to 0$, r\'ev\'elant ainsi que
l'instabilit\'e appara\^{\i}t en premier lieu pr\`es du cylindre
int\'erieur o\`u la courbure est plus importante. Les variations de la
fr\'equence et des composantes radiale et azimutale du vecteur d'onde
sont repr\'esent\'ees sur les Figs. 2a et 2b. Lorsque $T_1<T_2$, le
nombre d'onde azimutal s'annule pour une valeur de la courbure
$\Gamma^*=0.24$ alors qu'il reste fini lorsque $T_1>T_2$. Ces
r\'esultats sont en accord avec les observations exp\'erimentales
effectu\'ees par ombroscopie \cite{garnier00, nico}. 

\section{Conclusion}

Nous avons consid\'er\'e l'\'ecoulement qui appara\^{\i}t dans une mince
couche de fluide confin\'ee entre deux cylindres maintenus \`a des
temp\'eratures diff\'erentes. Loin des parois lat\'erales, l'\'etat de
base est mod\'elis\'e par un \'ecoulement lentement variable suivant la
direction radiale. Une analyse locale de stabilit\'e est effectu\'ee
conduisant aux valeurs critiques du nombre de Grashof, de la fr\'equence
et du vecteur d'onde en fonction de la courbure locale. Ces r\'esultats
montrent que l'instabilit\'e appara\^{\i}t en premier pr\`es du cylindre
int\'erieur avec un vecteur d'onde purement radial si ce cylindre est le
bord froid et dans le cas contraire ce sont des ondes spirales qui
apparaissent. Ces pr\'edictions th\'eoriques sont en tr\`es bon accord
avec les r\'esultats exp\'erimentaux expos\'es en \cite{garnier00},
\cite{nico}.

\newpage
  

\newpage

\newpage

\end{document}